\documentclass[aps,amsmath,amssymb,twocolumn,showpacs]{revtex4}

\usepackage{graphicx}

\begin{document}

\title{Saddles and dynamics in a solvable mean-field model}

\author{
        L. Angelani$^{1,2}$, G.~Ruocco$^{1}$, and F.~Zamponi$^{1}$
        }

\affiliation{
         $^1$Dipartimento di Fisica and INFM, Universit\`a di Roma
         {\em La Sapienza}, P. A. Moro 2, 00185 Roma, Italy\\
	$^2$SMC - INFM 
        Universit\`a di Roma {\em La Sapienza}, P. A. Moro 2, 00185 Roma, Italy
}
\date{\today}
\begin{abstract}
We use the saddle-approach, recently introduced in the numerical investigation of simple 
model liquids, in the analysis of a mean-field solvable system.
The investigated system is the $k$-trigonometric model,
a $k$-body interaction mean field system,
that generalizes the trigonometric model introduced 
by Madan and Keyes [J. Chem. Phys. {\bf 98}, 3342 (1993)]
and that has been recently introduced to investigate 
the relationship between thermodynamics and topology of the configuration space.
We find a close relationship between the properties of saddles 
(stationary points of the potential energy surface) 
visited by the system and the dynamics.
In particular the temperature dependence of saddle order follows that of the diffusivity,
both having an Arrhenius behavior at low temperature and a similar shape 
in the whole temperature range.
Our results confirm the general usefulness of the saddle-approach in the interpretation 
of dynamical processes taking place in interacting systems.
\end{abstract}
\pacs{31.50.-x, 61.20.Gy} 
\maketitle

\section{Introduction}
The {\it landscape paradigm} \cite{landscape}
is based on the idea that the relevant behavior of a given interacting system 
can be understood by studying
the general characteristics of the potential energy surface (PES),
i.e. the hypersurface defined by the potential energy as a function 
of the total configurational degrees of freedom.
This way of looking at the dynamics in liquids has received 
in the last few years new attention due to computational improvement and 
new theoretical approaches.
Within the large variety of studies based on the PES characteristics \cite{pes}
it has recently pointed out the relevant role played by 
stationary points (saddles) of the PES
in the analysis of the supercooled liquid regime of simple 
model liquids \cite{noi_sad,cav_sad,doye_wales,noi_jcp}.
The main result of the saddle-based approach has been the finding of
a close relationship between the parameters describing the 
structural arrest and the saddles visited at a given temperature.
More specifically, the saddle order
(fraction of negative curvatures -corresponding to negative eigenvalues of the Hessian- 
at the saddle points) 
turns out to have a well defined temperature dependence and extrapolates to zero 
at the mode-coupling temperature $T_{MCT}$ \cite{mct}, 
where the inverse diffusion coefficient and the structural relaxation time 
appear to diverge as a power law.
This result has been very useful to give a landscape-based interpretation  of the 
dynamic critical temperature for the model liquids investigated:
$T_{MCT}$ marks a crossover between a saddle-to-saddle dominated dynamics at ``high'' temperatures
and a minimum-to-minimum dominated dynamics below $T_{MCT}$.
Moreover, to further emphasize the role of the saddles in determining the slow dynamics in liquids, 
the temperature dependence of saddle order has been shown
to be quantitatively related to that of diffusivity \cite{noi_jcp},
allowing to give a simple interpretation of the relevant diffusive processes 
in terms of ``open'' directions in the $3N$ dimensional configuration space.
The analysis of saddles has also permitted to obtain important information about the shape 
of the landscape itself: it has been observed an high regularity in the distances 
among saddles and in their energy locations.
This suggested a surprising simple organization of those points of the PES, 
i.e. the saddles, that are relevant for the description of the slow dynamics.
The observed regularity allows us 
to introduce simplified models of the landscape reproducing some 
of the observed PES characteristics.
In a previous work \cite{noi_jcp} we explored the features of one of these simple models,
the trigonometric model (TM) introduced by Madan and Keyes \cite{madan_keyes}, 
made of $N$ independent sinusoidal degrees of freedom.
Despite its non-cooperative nature,
the trigonometric model seems to capture many important features of the PES of simple liquids.

In this work we use the saddle-approach to analyze another model system that generalizes the TM
introducing a cooperative interaction among the different degrees of freedom.
More specifically, in the model studied here -that we call $k$-trigonometric model ($k$TM)-
there is a $k$-body mean field interaction.
Our aim is to extend the validity of the relationship between 
landscape (saddle properties in our case) and dynamic behavior, 
besides the model liquids investigated before,
to systems where a cooperative behavior exists.
The choice of the $k$TM has been motivated for two main reasons:
it is a generalization of a simple liquid PES model, and  
it is a new model for testing in an analytical way the ideas underlying 
the landscape-dynamics relationship.
Indeed the thermodynamics of $k$TM, 
as the dynamics (diffusion coefficient), 
and the saddles visited at a given temperature 
are analytically computable.
This allows us to obtain directly a clear interpretation of the relationship between 
characteristics of the visited landscape and dynamics
in a truly cooperative system, avoiding the computational
problems usually encountered in the numerical analysis of PES in model liquids.
The main results are:
{(\it i)} we confirm the existence of a close relationship between saddle order and diffusivity:
both have a similar temperature behavior, and approach an asymptotic Arrhenius law
at low temperature;
{(\it ii)} the energy barrier parameter observed in the Arrhenius law of low temperature 
expansion of diffusivity is the mean elevation
$\delta e$ of saddles of order one from the minima, and this value grows linearly with the parameter
$k$ governing the number of interactions in the Hamiltonian.
Therefore, our findings confirm the generality of the relationship between landscape (saddle properties) and dynamics,
first observed numerically in simple liquids \cite{noi_sad,cav_sad,noi_jcp}
and analytically in spin glasses \cite{cav2}.

The paper is organized as follow:
in Section II we introduce the model and its characteristics;
in Section III the thermodynamics is exactly solved, using a mean-field effective Hamiltonian;
in Section IV we explore the properties of saddles sampled by the system at different temperatures;
in Section V the diffusion coefficient is determined through a Langevin dynamics,
and in Section VI we report the conclusions.

\section{The model}
The model we study is the mean-field $k$-trigonometric model ($k$TM), 
defined by the configurational Hamiltonian:
\begin{eqnarray}
\nonumber
H &=& \frac{\Delta}{N^{k-1}} \sum_{i_1,...,i_k} [1 -
\cos (\varphi_{i_1}+...+\varphi_{i_k})]  \\ 
&=& \frac{\Delta}{N^{k-1}} \sum_{i_1,...,i_k} [1 -
\Re ( e^{i\varphi_{i_1}} \cdots e^{i\varphi_{i_k}})] \ ,
\label{ktm}
\end{eqnarray}
and by the Langevin dynamics:
\begin{equation}
\gamma \dot{\varphi}_i = - \frac{\partial H}{\partial \varphi_i} + \eta_i \ ,
\label{langevin}
\end{equation}
where $\{\varphi_i\}$ are angular variables, $\varphi_i\!\in\! [0,2\pi)$,
$i\!=\!1,...,N$, $\gamma$ is the friction constant and $\eta_i$ is a Gaussian white noise
with $\langle \eta_i(t)\rangle\!=\!0$ and $\langle\eta_i(t)\eta_j(t')\rangle\!=\!2T\gamma\delta_{ij}\delta(t-t')$.
Defining the variable:
\begin{equation}
z(\{\varphi\}) = \frac{1}{N} \sum_i e^{i\varphi_i} =  \xi(\{\varphi\})\  e^{i \psi(\{\varphi\})}  \ , 
\end{equation}
the Hamiltonian of the $k$TM can be written in a more compact way:
\begin{equation}
H = N \Delta [ 1 - \Re (z^k)]  = N \Delta ( 1 - \xi^k \cos k \psi) \ .
\label{ktm_1}
\end{equation}
The model is invariant under the discrete group $C_{kv}$, generated by the transformations:
$\varphi_i  \rightarrow \varphi_i + 2\pi/k$ and $\varphi_i \rightarrow -\varphi_i$.
It can be considered as a model for a system of bidimensional Heisenberg 
spins with an interaction that is not invariant under the rotation group but only under the
$C_{kv}$ group.
Depending on the value of the parameter $k$ governing the number of interactions, the system 
manifests a wide spectrum of thermodynamic behavior \cite{acprz}:
\begin{itemize}
\item 
For $k\!=\!1$ the $k$TM reduces to the trigonometric model introduced by Madan and Keyes \cite{madan_keyes}
(the relation between saddles and diffusivity for $k\!=\!1$ has been investigated in Ref. \cite{noi_jcp}):
the system reduces to independent degrees of freedom and
all the thermodynamic quantities are smooth functions of temperature.
\item
For $k\!=\!2$ a second order phase transition takes place, with a
spontaneous breaking of the $C_{kv}$ symmetry and a spontaneous magnetization.
\item
For $k\!\geq\!3$ the system undergoes a first order phase transition, again with a 
magnetization breaking the symmetry group.
\end{itemize}
While in a previous paper we have analyzed the relationship between 
phase transition points and topological changes in the landscape 
for this model \cite{acprz},
here we are interested in the investigation of a link between characteristic of the landscape 
explored by the system at a given temperature and its dynamic behavior.
As the high temperature paramagnetic phase is quite trivial (the effective potential energy is constant 
and the system can be viewed as an ensemble of free particles), 
the interesting region for our purpose will be the low $T$ one, 
in which the magnetization is different from zero.

\begin{figure}[t]
\centering
\includegraphics[width=.47\textwidth,angle=0]{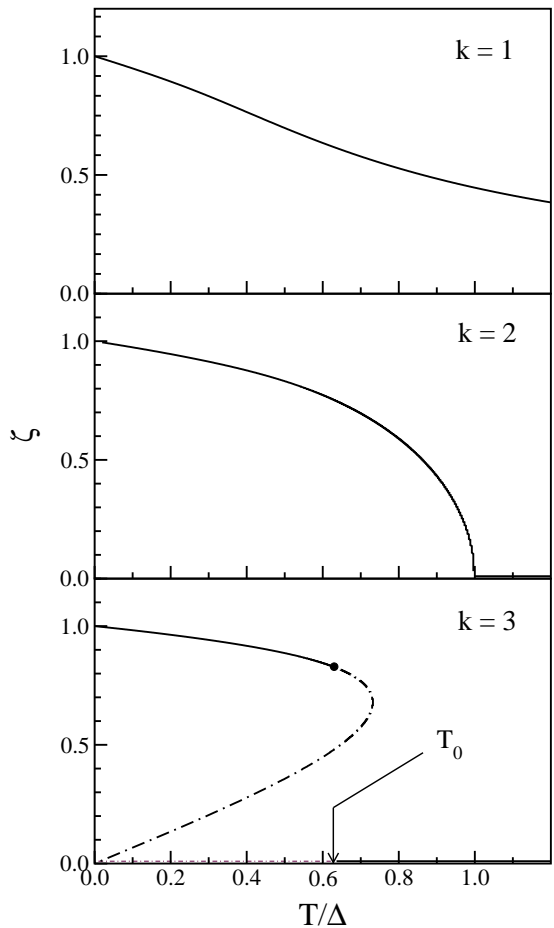}
\caption{
Mean magnetization $\zeta$ as a function of temperature for $k\!=\!1,2,3$.
For $k\!=\!3$ the value of the canonical transition temperature $T_0\!=\!0.63$ is indicated by an arrow,
and dot-dashed lines represent the canonical thermodynamical inaccesible region 
(the same representation is used in all the figures).
}
\label{fig1}
\end{figure}

\section{Thermodynamics}
The thermodynamics of the $k$TM is exactly solvable.
In a previous paper \cite{acprz} we 
reported the exact derivation of the partition function using 
the saddle point technique. 
Here we report a different and more simple derivation based on the   
introduction of an effective Hamiltonian for each single degree of freedom.
Using the standard approach,
in order to obtain a mean-field form of the Hamiltonian,
we have to make the substitution in Eq. \ref{ktm}
\begin{eqnarray}
e^{i\varphi_{i_1}} \cdot \cdot \cdot e^{i\varphi_{i_k}}
&\rightarrow& k \ e^{i\varphi_{i_1}}  \langle e^{i\varphi_{i_2}}\rangle  
\cdot \cdot \cdot \langle e^{i\varphi_{i_k}} \rangle 
\nonumber \\
&-& (k-1) \ \langle e^{i\varphi_{i_1}}\rangle  \cdot \cdot \cdot \langle e^{i\varphi_{i_k}} \rangle \ .
\end{eqnarray}
We introduce the parameter $\zeta\!=\!\langle e^{i\varphi}\rangle$,
that has to be determined self-consistently; 
the mean-field Hamiltonian is then:
\begin{equation}
{\cal H} = \Delta [ 1 + (k-1) \zeta^{k} - k \zeta^{k-1} \cos \varphi ] \ ,
\label{Heff}
\end{equation}
where we have chosen $\zeta$ to be real, 
without loss of generality (this corresponds to choose a 
particular magnetization of the low temperature state).
In the canonical ensemble the self consistency equation for $\zeta$ turns out to be:
\begin{equation} 
\zeta  = \langle \cos \varphi \rangle _{\cal H} =
\frac{I_1 (\beta \Delta k \zeta^{k-1})}{I_0 (\beta \Delta k \zeta^{k-1})} \ ,
\label{zeta}
\end{equation}
where $\beta\!=\!(K_B T)^{-1}$,    
$I_0 (\alpha) = (2\pi)^{-1}\int_0^{2\pi} d\varphi e^{\alpha \cos \varphi}$ and
$I_1 (\alpha) = I_0^{'} (\alpha)$
are the modified Bessel functions of order 0 and 1 respectively.
\begin{figure}[t]
\centering
\includegraphics[width=.47\textwidth,angle=0]{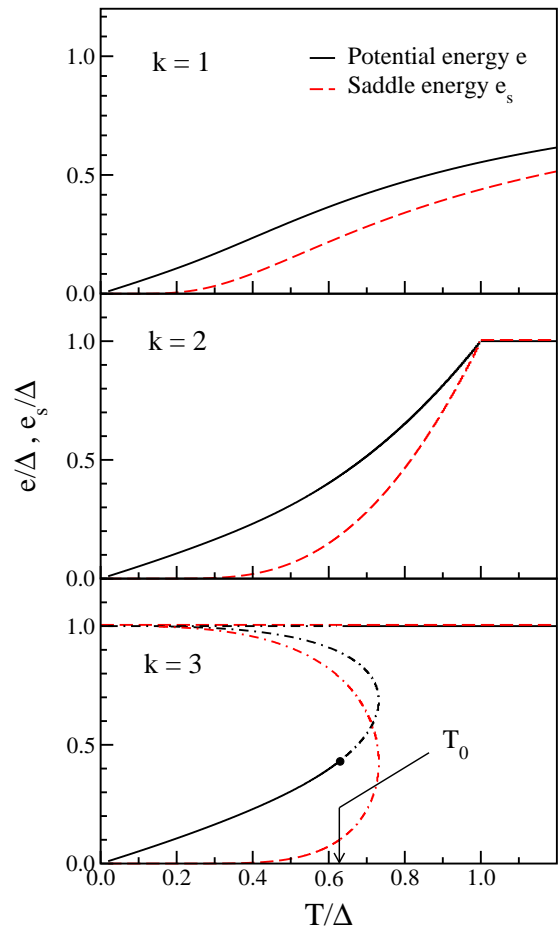}
\caption{Potential energy $e$ (full lines) and saddle energy $e_s$ (dashed lines) as a function of temperature
for $k\!=\!1,2,3$.
$T_0\!=\!0.63$ is the canonical transition temperature for $k\!=\!3$,
and dot-dashed lines represent the canonical thermodynamical inaccesible region.
}
\label{fig2}
\end{figure}

For each $\beta$ the Eq. \ref{zeta} gives the thermodynamic value of the mean magnetization 
$\zeta (\beta)$. 
The value $\zeta\!=\!0$ always solve Eq. \ref{zeta}, but is a stable 
solution only at low $\beta$ (high temperature).
As $\beta$ is increased others solutions appear at 
a given $\beta$, and the one with lower free energy is the stable one.
In Fig. \ref{fig1} we report the function $\zeta(\beta)$ for $k\!=\!1,2,3$.
For $k\!=\!1$ the curve is smooth, as no phase transition occurs.
For $k\!=\!2$ there is a singular point at which a second order phase transition takes place, 
separating an high temperature paramagnetic phase ($\zeta\!=\!0$) 
and a low temperature ordered phase, with two possible 
magnetizations corresponding to $\psi\!=\!0$ and $\psi\!=\!\pi$.
For $k\!=\!3$ (and also for $k\!>\!3$, not reported in the figure) the system undergoes a 
first order phase transition, corresponding to the appearance of three solutions 
of Eq. \ref{zeta}: the paramagnetic one $\zeta\!=\!0$ (always present), 
the magnetic one $\zeta>0$ 
(with degeneration $k$, i.e. $k$ different possible value of $\psi$) 
and that corresponding to the maximum of the free energy separating the two previous solutions.
For the $k\!=\!3$ case we report all the solutions of Eq. \ref{zeta} at different temperatures:
the three magnetization values at a given low temperature then correspond to
these three  different solutions.
We report with an arrow in the figure the transition temperature $T_0\!=\!0.63$,
marking the transition point at which the $k\!=\!3$ model exhibits a first order 
phase transition (in the figure, as well as in all the other figures,
the canonical thermodynamically inaccessible region is represented by 
dot-dashed lines).

We can write the canonical probability distribution for the variable $\varphi$ 
at a given temperature in the form:
\begin{equation}
{\cal P}(\varphi;\beta) = \frac{e^{-\beta {\cal H}}}{\cal Z} = 
\frac{e^{\beta \Delta k \zeta^{k-1} \cos \varphi}}
{2\pi I_0(\beta \Delta k \zeta^{k-1})} \ , 
\label{probphi}
\end{equation}
where ${\cal Z}$ is the canonical partition function for the effective Hamiltonian ${\cal H}$ and 
the parameter $\zeta (\beta)$ is a function of temperature as obtained from Eq. \ref{zeta}.

Using Eq. \ref{probphi} it is easy to calculate the different static quantities we are interested in.
For example, the mean potential energy $e(\beta)$ results to be:
\begin{equation} 
e(\beta) = \Delta ( 1 - \zeta^k ) \ , 
\end{equation} 
as one can also see directly substituting the mean magnetization value $\zeta$ in Eq. \ref{ktm_1}.
In Fig. \ref{fig2} the temperature dependence of potential energy $e$ (full lines in the figures) 
is shown for $k\!=\!1,2,3$. 
For a discussion of the different behaviors see the comments of Fig. \ref{fig1} reported above.
 
\begin{figure}[t]
\centering
\includegraphics[width=.47\textwidth,angle=0]{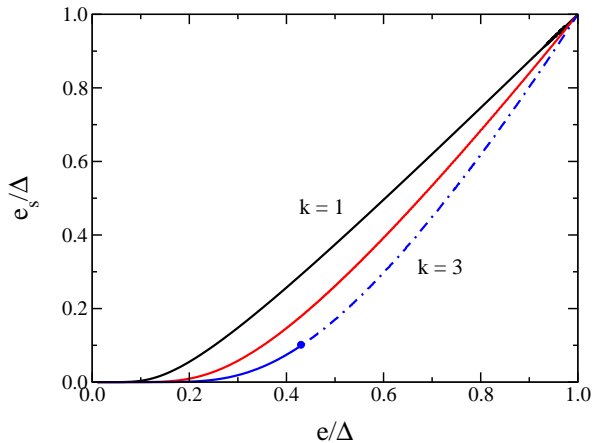}
\caption{Saddle energy $e_s$ against potential energy $e$ for $k\!=\!1,2,3$.}
\label{fig3}
\end{figure}

\section{Saddles}
The saddles are the stationary points of the potential energy,
satisfying the zero force equation: $\nabla H \!=\! 0$. 
We are interested in locating the saddles of energy $e_s$ and 
in finding their order $n$.
The latter quantity is defined as the fractional number of negative eigenvalues of the Hessian
matrix calculated at the saddle point, 
i.e. the fractional number of downward curvature directions.
For the $k$TM it turns out that the saddles located at a given energy level all
have the same order $n$, and 
the relationship between $e_s$ and $n$ is analytical (see Ref. \cite{acprz}):
\begin{equation}
e_s(n) = \Delta [1 - (1-2n)^k]  \ ,
\label{e_sad}
\end{equation}
valid for $n<1/2$ (or $e<1$), 
condition satisfied in the whole thermodynamically accessible phase space.
Similarly to the case of liquids \cite{noi_sad,cav_sad}, 
we introduce the concept of saddles visited 
during the dynamical evolution. To this aim we define a partition of the configurational
space, such that each point in $\mathbb{R}^N$ is associated to a saddle point 
(i.e. we define a basin of attraction for a saddle).
The partition of $\mathbb{R}^N$ in basins of saddles of $H$ is realized by introducing the pseudo-potential
$W = |\nabla H|^2$ and identifying the basin of attraction of the minima of $W$ with that of a saddle
of $H$. Indeed, $W$ has the property that absolute minima correspond 
to saddles of $H$ \cite{nota_w}.
As for the thermodynamic calculation, we can use also in this case 
an ``effective'' pseudo-potential ${\cal{W}}$ given by:
\begin{equation}
{\cal{W}}  \simeq \left|\frac{\partial {\cal H}}{\partial \varphi}\right|^2 =
\Delta^2 k^2 \zeta^{2(k-1)} \sin^2\varphi \ .
\label{pseudopot}
\end{equation}
The fact that the ``effective'' pseudo-potential can be obtained differentiating the
effective Hamiltonian is not obvious ``a priori''.
In ref. \cite{zampo} we will show that Eq. \ref{pseudopot} 
is correct at small enough temperature, that is the temperature 
region we are interested in.
The search of the minima of $W$ goes through a minimization procedure,
i.e. through the solution of the steepest descent equation of the form:
\begin{equation}
\left\{ 
\begin{array}{ll}
\dot{\varphi} = - \nabla {\cal{W}} = - \Delta^2 k^2 \zeta(t)^{2(k-1)} \sin 2\varphi \ , \\
\zeta(t) = \langle \cos \varphi(t) \rangle \ , \\
\end{array}
\right. 
\label{steep}
\end{equation}
where now the parameter $\zeta$ is a function of time $\zeta(t)$,
determined by a self consistency dynamic equation (in the above expression 
$\langle \cdot \rangle$ means average over initial conditions).
We are interested in the infinite time solutions of Eq. \ref{steep},
$\varphi(t\!\rightarrow\!\infty|\varphi_0)$,
as a function of initial conditions 
$\varphi_0 \equiv \varphi(t\!=\!0)$.
Without explicitly solving Eq. \ref{steep}, 
we observe that the sign of $\nabla {\cal{W}}$ at fixed 
$\varphi$ does not change during time, due to the fact that the time dependent factor 
in Eq. \ref{steep}
is always positive.

\begin{figure}[t]
\centering
\includegraphics[width=.47\textwidth,angle=0]{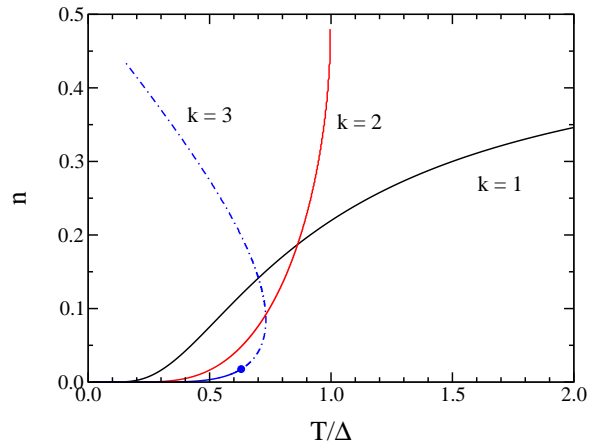}
\caption{Saddle order $n$ as a function of temperature $T$ for $k\!=\!1,2,3$.}
\label{fig4}
\end{figure}
This implies that the specific time dependence of $\zeta$ does not affect the final point 
$\varphi(t\!\rightarrow\!\infty|\varphi_0)$
reached from a given initial condition,
rather it controls the rapidity of approaching the final point. 
It it easy to see that the solutions are the following:
\begin{equation}
\varphi(t\!\rightarrow\!\infty|\varphi_0) = 
\left\{
\begin{array}{ll}
0 & \hspace{.5cm} \mbox{ if \ \  $\varphi_0 \notin (\frac{\pi}{2},\frac{3\pi}{2})$ } \ , \\ \\
\pi & \hspace{.5cm} \mbox{ if \ \  $\varphi_0 \in (\frac{\pi}{2},\frac{3\pi}{2})$ } \ .
\end{array}
\right.
\end{equation}
The solution $\varphi \!=\!0$ is a minimum of ${\cal{W}}$ corresponding to 
a minimum of the effective potential energy (Eq. \ref{Heff}), 
while the solution $\varphi \!=\!\pi$ is a minimum of ${\cal{W}}$ that 
corresponds to a maximum of the effective potential.
The mean saddle order $n$ as a function of temperature is then obtained as the probability to
be in the basin of attraction (defined through the pseudo-potential ${\cal{W}}$) 
of the maximum of the potential energy, or, in other words,
weighting the function 
$\delta (\varphi(t\!\rightarrow\!\infty|\varphi_0) - \pi)$
over the equilibrium initial conditions:
\begin{eqnarray}
\nonumber
n(\beta) &=& \int_0^{2\pi}d\varphi_0 \ {\cal P}(\varphi_0;\beta) \ \delta (\varphi(t\!\rightarrow\!\infty|\varphi_0) - \pi) \\
&=& \int_{\pi/2}^{3\pi/2}d\varphi_0 \ {\cal P}(\varphi_0;\beta) \ .
\end{eqnarray}
Using Eq. \ref{probphi} we finally obtain
\begin{equation}
n(\beta) = \frac{1}{2} \left[ 1 - \frac{L_0(\beta \Delta k \zeta^{k-1})}{I_0(\beta \Delta k \zeta^{k-1})}
\right] \ ,
\label{ord_sad}
\end{equation}
where we have introduced the modified Struve function of order $0$:
$L_0 (\alpha) = 2 \pi^{-1} \int_0^{\pi/2}  d\varphi \sinh (\alpha \cos \varphi)$.
The energy of saddles sampled by the system at different temperatures is then obtained 
using Eq. \ref{e_sad}, in which the temperature dependence is in 
the order $n(\beta)$, as expressed in Eq. \ref{ord_sad}: $e_s(\beta) = e_s [n(\beta)]$.
\begin{figure}[t]
\centering
\includegraphics[width=.47\textwidth,angle=0]{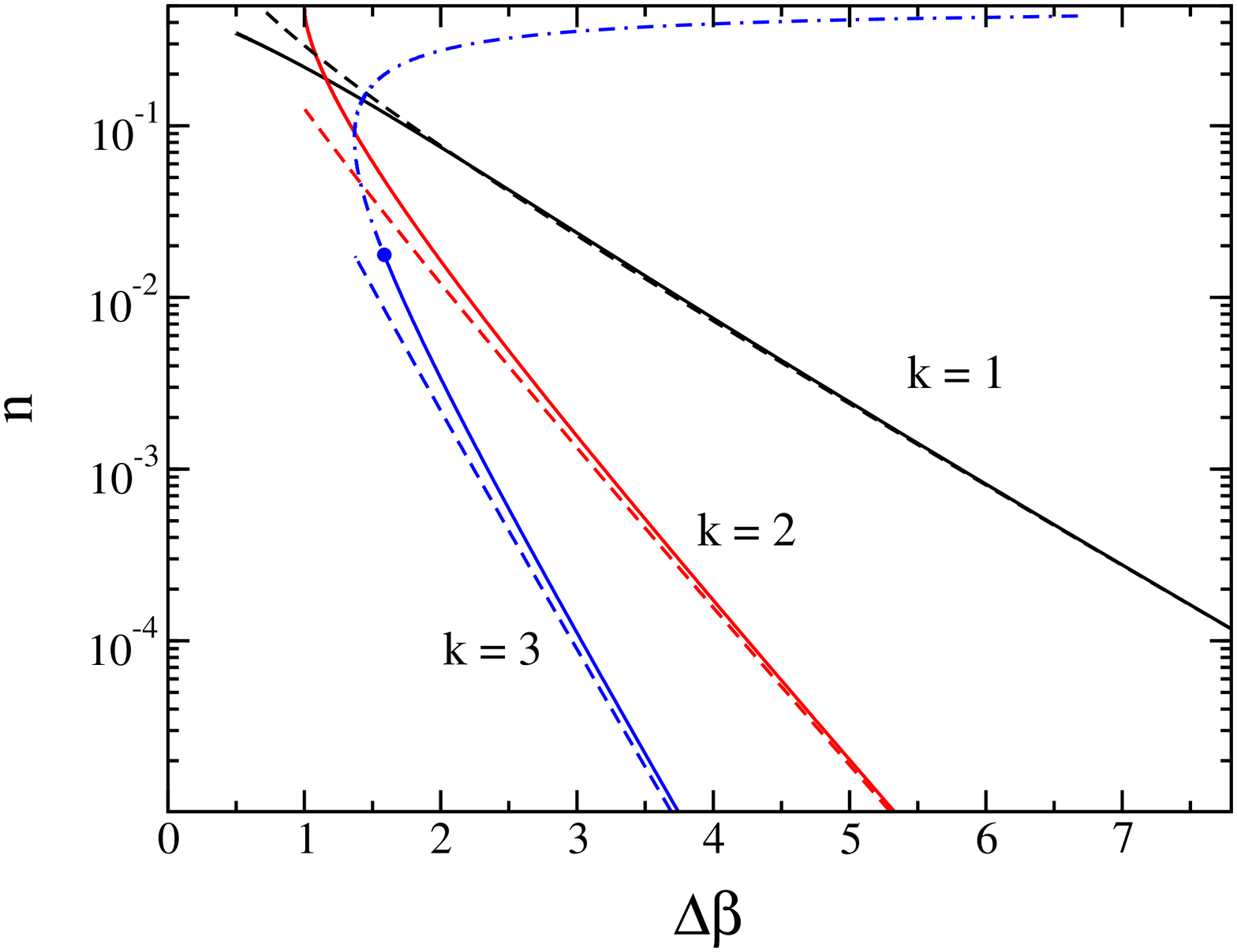}
\caption{
Saddle order $n$ as a function of inverse temperature $\beta$
for $k\!=\!1,2,3$ in a semilogarithmic plot;
dashed lines are the low temperature approximations, given by Eq. \ref{ord_sad_low}.
}
\label{fig5}
\end{figure}

In Fig. \ref{fig2} the saddle energies $e_s$ (dashed lines) are reported as a function of temperature
for $k\!=\!1,2,3$: qualitatively $e_s$ reproduces the shape of the potential energy $e$, 
and it is always below $e$, but coincident in the paramagnetic 
region for $k\!\geq\!2$.
The map $e_s$ vs. $e$ is shown in Fig. \ref{fig3}, where one observes that, growing the parameter $k$,
the energy difference between instantaneous configurations and saddles becomes more and more pronounced.

In Fig. \ref{fig4} the saddle order $n$ is reported as a function of temperature for $k\!=\!1,2,3$.
For $k\!=\!3$ the negative slope branch corresponds to the similar branch in Fig. 1, 
that is the solution of Eq. \ref{zeta} which is a maximum of the free energy
(in the figure we do not report the paramagnetic branch).
In order to visualize better the low temperature regime, in Fig. \ref{fig5} we plot the saddle order $n$
in a semilogarithmic scale as a function of inverse temperature $\beta$.
Dashed lines in the figure are the low temperature (high $\beta$) approximations
to Eq. \ref{ord_sad}:
\begin{equation}
n(\beta \gg 1) \simeq \sqrt{\frac{2 e^{k-1}}{\pi k \beta \Delta}} e^{-\beta \Delta k} \ ,
\label{ord_sad_low}
\end{equation}
which corresponds to an Arrhenius behavior.
\begin{figure}[t]
\centering
\includegraphics[width=.47\textwidth,angle=0]{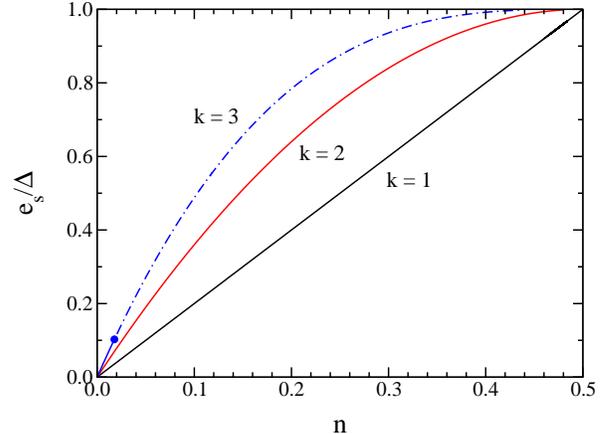}
\caption{Saddle energy $e_s$ against saddle order $n$ for $k\!=\!1,2,3$.}
\label{fig6}
\end{figure}

An important characteristic of saddles is the relationship between energy and order,
which is a feature of the energy landscape itself, independent on temperature.
In Fig. \ref{fig6} we report $e_s$ vs. $n$, as expressed by Eq. \ref{e_sad}.
The slope of the curves at $n\!=\!0$ gives information about the energy elevation 
$\delta e$ of saddles of order 1 from underlying minima (in this model all
the minima have zero energy), 
an interesting quantity in the analysis of activated processes in simple liquids.
For the $k$TM, using Eq. \ref{e_sad}, we have
\begin{equation}
\delta e = \left. \frac{de_s}{dn} \right|_{n=0} = 2\Delta k \ ,
\label{barrier}
\end{equation}
indicating an energy barrier value growing linearly with the parameter $k$.
\begin{figure}[t]
\centering
\includegraphics[width=.47\textwidth,angle=0]{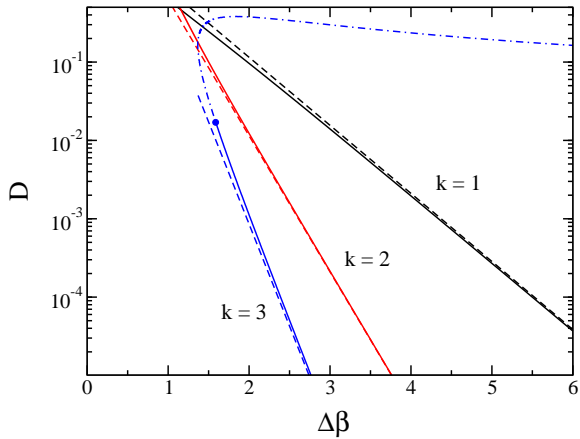}
\caption{Diffusivity $D$ as a function of inverse temperature $\beta$ for $k\!=\!1,2,3$.
Dashed lines are the low temperature approximations, given by Eq. \ref{diff_low}.}
\label{fig7}
\end{figure}

\section{Diffusivity}
As we want to establish a connection between landscape and dynamics,
we now turn our attention to the dynamics of the $k$TM.
For simple liquids a very interesting relation has been established 
between saddles and diffusion coefficient, so, also in our case, we 
focus on this dynamical quantity.
In order to determine the diffusivity $D$, 
we calculate the mobility in linear response.
We can use again the effective Hamiltonian Eq. \ref{Heff}, 
as one can show \cite{zampo}, using standard techniques \cite{lc}, 
that the single particle Langevin dynamics is the same if computed 
with the full Hamiltonian $H$ (Eq. \ref{ktm}) or 
the effective Hamiltonian ${\cal H}$ (Eq. \ref{Heff}).
The final result is then \cite{risken} :
\begin{equation}
D(\beta) = 
(\gamma \beta)^{-1} 
[ I_0 (\beta \Delta k \zeta^{k-1}) ]^{-2} \ ,
\label{diff}
\end{equation}
where $\gamma$ is the friction in the Langevin equation.

In Fig. \ref{fig7} the diffusion coefficient $D$ (Eq. \ref{diff}) is reported 
as a function of inverse temperature $\beta$ in a semilogarithmic scale,
together with the (Arrhenius) low temperature expansion of Eq. \ref{diff}:
\begin{equation}
D(\beta \gg 1) \simeq \frac{2\pi \Delta k e^{k-1}}{\gamma} e^{-\beta 2 \Delta k} \ .
\label{diff_low}
\end{equation}
\begin{figure}[t]
\centering
\includegraphics[width=.47\textwidth,angle=0]{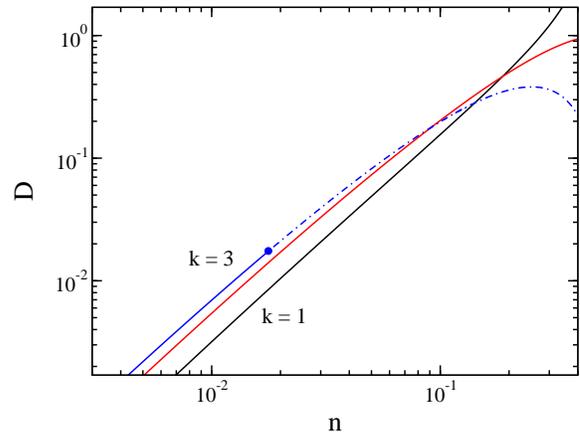}
\caption{Diffusivity $D$ against saddle order $n$ in double-logarithmic  scale for $k\!=\!1,2,3$.}
\label{fig8}
\end{figure}
It is interesting to note that the energy barrier value $\delta e$ in the Arrhenius low temperature 
expansion is exactly the energy elevation of saddles of order $1$ from minima (see Eq. \ref{barrier})
$\delta e = 2\Delta k$, as expected in the presence of activated processes.
  
For completeness in Fig. \ref{fig8} we report the diffusivity $D$ against the saddle order $n$ in a logarithmic scale.
From this plot one can conclude that the slow diffusive dynamics is 
controlled by the order of the saddles visited at each temperature.

\section{Conclusions}
In conclusion we have studied an exactly solvable mean-field model, the $k$TM, 
that generalizes the non-interacting trigonometric model introduced in Ref. \cite{madan_keyes}.
The useful framework of landscape saddle-approach, introduced in the numerical investigation of 
simple liquids, has been applied in our case,
evidencing how the relation between sampled saddles and dynamics works well 
also in our model: diffusion coefficient and saddle order have a very similar 
temperature dependence, both approaching an Arrhenius law at low temperature
and having a similar shape in the whole temperature range. 
The analysis of the $k$TM seems to evidence a quite general relationship 
between dynamics and saddles visited in the potential energy surface,
beyond the simple liquid models in which it was firstly observed.

\begin{acknowledgments}
We thank  L.~Cugliandolo and J.~Kurchan for pointing out the analytical techniques described in \cite{lc} 
and for very useful and interesting discussions. We also thank F.~Sciortino for carefully reading the manuscript.
We acknowledge financial support from INFM
and MURST COFIN2000.
\end{acknowledgments}

\clearpage
\newpage


\begin{thebibliography}{99}
\bibitem{nota_w}
We note that in this case all the found minima of $W$ are true saddles of $H$,
in opposite to the Lennard-Jones case, in which the majority of minima of $W$ are local minima 
corresponding to inflection directions \cite{doye_wales,noi_jcp}.
\end{thebibliography}
\end{document}